\documentclass[10pt,letterpaper]{article}
\usepackage{opex3}

\begin{document}

\title{Reconstructing an Icosahedral Virus from Single-Particle
Diffraction Experiments}

\author{D. K. Saldin, H.-C. Poon, P. Schwander, M. Uddin, and M. Schmidt}

\address{Department of Physics, University of Wisconsin-Milwaukee, P. O. Box 413, \\ Milwaukee, WI 53201}

\email{dksaldin@uwm.edu} 



\begin{abstract}

The first experimental data from single-particle scattering experiments
from free electron lasers (FELs) are now becoming available. The first such
experiments are being performed on relatively large objects such as
viruses, which produce relatively low-resolution, low-noise diffraction 
patterns in so-called ``diffract-and-destroy'' experiments.
We describe a very simple test on the angular correlations of measured
diffraction data to determine if the scattering is from an icosahedral
particle. If this is confirmed, the efficient algorithm proposed can then
combine diffraction data from multiple shots of particles in random 
unknown orientations to generate
a full 3D image of the icosahedral particle. We demonstrate this
with a simulation for the satellite tobacco necrosis virus (STNV), the
atomic coordinates of whose asymmetric unit is given in Protein
Data Bank entry 2BUK. 
\\
\end{abstract}


\ocis{(100.3200) Image Processing: Inverse Scattering;
(190.5825) Scattering: Scattering Theory.} 


\section{Introduction}

The free electron lasers (FELs) now beginning to come online produce radiation
many orders of magnitude brighter than than any existing source, and
enable experiments previously the domain only of science fiction. One
such proposed experiment \cite{neutze2000} envisages reconstructing the 
3D structure of a microscopic entity such as a virus from many ultrashort 
diffraction patterns of many identical copies of the
particles in random orientations from single pulses of FEL radiation. 
Although the particles will undoubtedly suffer catastrophic radiation
damage, the ultrashort nature of FEL radiation is expected to produce
diffraction patterns of the particles before significant disintegration.
An experiment on individual mimivirus particles was reported recently
\cite{seubert2011}. The paper illustrates convincing diffraction patterns
of the virus particle in two different orientations, from which 2D
projections of the particles are reconstructed using an iterative phasing
algorithm. Although such particles are known to be largely icosahedral,
little evidence of the icosahedral shape is evident in the reconstructed
projections. Several algorithms have been proposed for
reconstructing a full 3D image of the particle 
from an ensemble of many such diffraction patterns from
randomly oriented particles. The methodology followed by some of these
approaches \cite{shneerson2008,fung2009,loh2009} is to find the likely
orientation of the measured diffraction patterns in the 3D reciprocal
space of the particle. 

Another approach \cite{saldin2009} dispenses with finding the likely 
orientations of the individual diffraction patterns by integrating over 
orientations, in an attempt to find the spherical harmonic representation of 
the 3D diffraction volume of a single particle from the averages of the angular
correlations of the intensities on the measured diffraction patterns. This
method of analysis is even applicable to individual diffraction patterns  
from multiple identical particles \cite{kam1977}. The particles need to be 
frozen in space or time while the scattering is taking place. If the 
scattering is from a single FEL pulse of radiation, the particles will be 
essentially frozen in time for the duration of the scattering even if not 
frozen in space. This opens this method to the analysis of scattering from 
particles in random orientations within a droplet. With such an approach, the
``hit rate'' in an experiment with a FEL can become 100\%, whereas a low 
hit rate is to be expected when attempting to hit submicron particles with a
submicron pulsed laser beam. The signal-to-noise ratio from such
snapshot patterns is independent of the number of particles per
shot, but increases with the square root of the number of shots
\cite{kirian2011}.
This approach also has the advantage that it operates on a compressed
version of the voluminous data produced by a FEL. 

We point out here another advantage of this approach: it is easily 
amenable to simplifications resulting from any known point-group symmetry
of the particles under study.  This is a powerful advantage for the study
of virus structure, which is dominated by that of its protein coat which 
encloses the genetic material, DNA or RNA, which contain the instructions
for the replication of the virus. To quote from Caspar and Klug 
\cite{caspar1962}  
``there are only a limited number of efficient 
designs possible for a biological container which can be constructed 
from a large number of identical protein molecules, The two basic designs
are helical tubes and icosahedral shells''. 
Viruses have regular shapes
since they are formed by the self assembly of identical protein subunits
which are coded by the limited quantity of genetic material capable of being
stored within the small volume enclosed by its protein coat. An icosahedron, 
for example can be formed by the self assembly of at least 60 identical 
subunits.  The genetic material needs to code for just one of these subunits, 
a factor of at least 60 smaller than the entire structure.

\section{Icosahedral Harmonics}

The first aim of this approach is to find the spherical harmonic representation
of the intensity distribution of any resolution shell in the reciprocal
space of a single particle. Any prior information about the nature of
this distribution may be incorporated by limiting the set of spherical
harmonics over which the summation is performed and by any relationship
amongst the amplitudes of the different spherical harmonics which are
a consequence of any known point-group symmetry. 

An obvious restriction of the form of the intensity distribution 
\begin{equation}
I(q,\theta,\phi)=\sum_{lm} I_{lm}(q) Y_l^m(\theta,\phi)
\label{genexp}
\end{equation}
is its known inversion (or Friedel) symmetry. Since
\begin{equation}
Y_{l}^m(\pi-\theta,-\pi+\phi) = (-1)^l Y_{l}^m(\theta,\phi)
\end{equation}
it follows that a spherical harmonic expansion of an intensity
distribution may contain only even values of the angular momentum quantum
number $l$. The fact that the intensity distribution is real, allows
the restriction to a summation over just the so-called real spherical
harmonics (RSHs) $S_{l}^{m}(\theta,\phi)$ defined by the combinations of
spherical harmonics: 
\begin{equation}
S_{l}^{m}(\theta,\phi) = \left\{ 
\begin{array}{ll}
\frac{1}{\sqrt{2}} 
\left[ Y_{l}^{m}(\theta,\phi)  + (-1)^m Y_l^{-m}(\theta,\phi) \right] & m>0 \\
Y_{l}^{0}(\theta,\phi) & m=0 \\
\frac{1}{i\sqrt{2}}
\left[ Y_l^m(\theta,\phi) - (-1)^m Y_l^{-m}(\theta,\phi) \right] & m< 0
\end{array}
\right.
\end{equation}
where the set of RSH's with $m\ge 0$ form a set,
whose $\phi$ dependence is of the form $\cos{(m\phi)}$, and the set
with $m\le 0$ likewise a set with $\phi$ dependence
of the form $\sin{(m\phi)}$. If the reconstructed
intensity distribution has a mirror plane, this may be chosen to be the
$x-z$ plane, or the plane for which $\phi=0$. Then (\ref{genexp}) 
may be replaced by a summation over only the subset of RSHs for which 
$m\ge 0$, and we may take
\begin{equation}
I(q,\theta,\phi)=\sum_{l,m\ge 0} R_{lm}(q) S_l^m(\theta,\phi).
\label{realexp}
\end{equation}
Since both the right hand side (RHS) and the left hand side (LHS) of the above 
equation are real, the coefficients $R_{lm}(q)$ may also be taken as real.

\begin{figure}[htbp]
\centering\includegraphics[width=14cm]{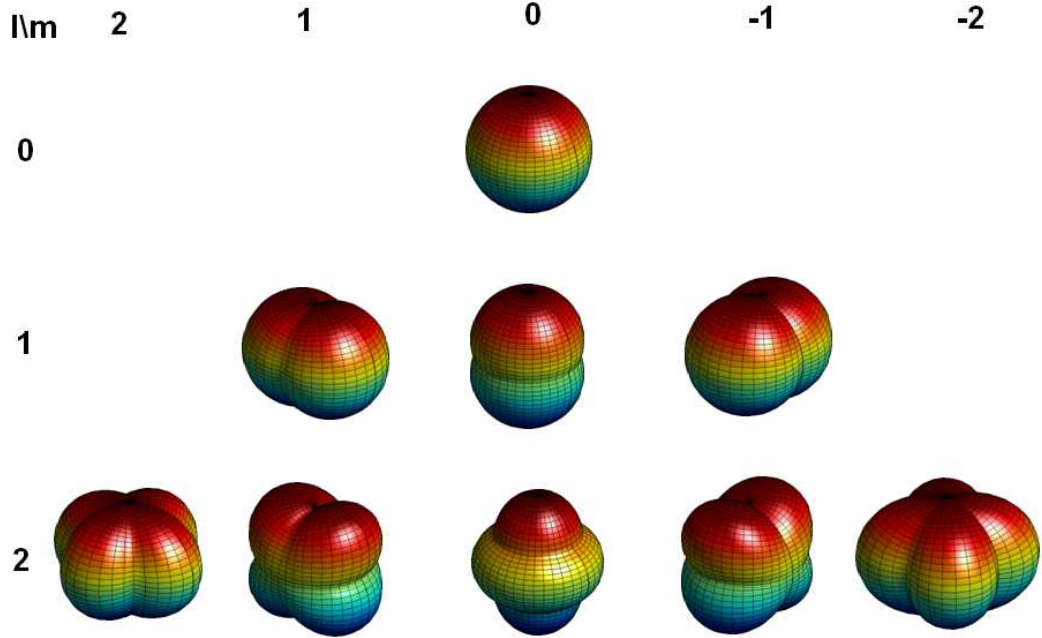}
\caption{Visualization of real spherical harmonics (RSHs) of angular momentum 
quantum numbers $l=$ 0,1, and 2. The plots are made with MATLAB by an 
adaptation of software by Denise L. Chan (avilable from the Mathworks 
web site).} 
\label{sphrm}
\end{figure}

\begin{figure}[htbp]
\centering\includegraphics[width=12cm]{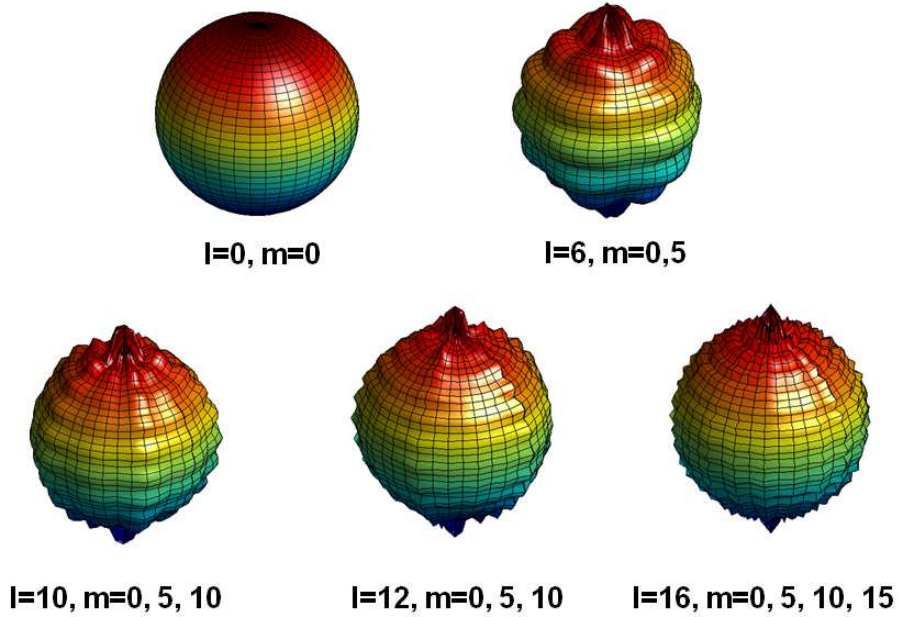}
\caption{Icosahedral harmonics of angular momentum quantum number $l=$ 
0,6,10,12 and 16. Each is a linear combination of the RSH's of the magnetic
quantum numbers indicated. Visualization by the same software as
Fig. \ref{sphrm}} 
\label{icoharm}
\end{figure}

The 3D polar plots of Fig. \ref{sphrm} display the familiar forms of the 
RSHs for the values of $l=$ 0,1,2.
Further point group symmetries of $I(q,\theta,\phi)$ result in still
further restrictions on allowed terms of the general expansion
(\ref{genexp}) above. When the 3D intensity distribution has icosahedral
symmetry, for example, (\ref{genexp}) may be replaced by
\begin{equation}
I(q,\theta,\phi)=\sum_l g_l(q) J_l(\theta,\phi),
\label{icoexp}
\end{equation}
where the quantities $J_l(\theta,\phi)$ are known as
icosahedral harmonics (IHs), specified up to and including $l=$30 by
only the angular momentum quantum number $l$. Since the orientation of our
reconstructed 3D intensity distribution in the frame of reference of the
particle may be chosen arbitrarily, the $x-z$ plane may be chosen to be the
mirror plane, allowing the IHs 
(\ref{icoexp}) to be constructed from just the RSHs of positive $m$, i.e. 
\begin{equation}
J_l(\theta,\phi) = \sum_{m\ge 0} a_{lm} S_l^m(\theta,\phi)
\label{icodef}
\end{equation}
where the coefficients $a_{lm}$ are the real numbers for normalized
RSHs tabulated by e.g. Jack
and Harrison (1975) \cite{jack1975}, the ones for the lowest
allowed even values of $l$ being reproduced in Table \ref{table1}. 
Since the IHs
involve a sum over the magnetic quantum number, at least up to $l=30$,
they depend on the quantum number $l$ only. The forms of the icosahedral
harmonics of lowest even degree, $l$=0,6,10,12, and 16 are illustrated in 
Fig. \ref{icoharm}
using the same 3D polar plots.

\begin{table}[c]
\caption{Expansion coefficients of the lowest even degree icosahedral harmonics
with z-axis chosen to be the 5-fold rotation axis.
For the list up to $l$=30, see e.g.  \cite{jack1975}. Note the rows are
characterized by $l$ and the columns by $m$.}
\vspace{5mm}
\begin{tabular}{|c|c|c|c|c|c|}
\hline
l$\backslash$m & 0 & 5 & 10 & 15 & 20 \\ 
\hline
0 & 1.0 & & & & \\
\hline
6 & 0.531085 & 0.847318 & & & \\
\hline
10 & 0.265539 & -0.846143 & 0.462094 & & \\
\hline
12 & 0.454749 & 0.469992 & 0.756513 & & \\
\hline
16 & 0.334300 & -0.493693 & -0.634406 & 0.491975 & \\
\hline
18 & 0.399497 & 0.450611 & 0.360958 & 0.712083 & \\
\hline
20 & 0.077539 & -0.460748 & 0.747888 & -0.231074 & 0.411056 \\
\hline
\end{tabular}
\label{table1}
\end{table}

Note that since the RSH's $S_l^m(\theta,\phi)$ are orthonormal with
respect to integrations over spherical shell, the icosahedral
harmonics $J_l(\theta,\phi)$ will also be orthonormal with respect to
the same integration provided
\begin{equation}
\sum_m a^2_{lm } = 1, \ \ \ \ \ \forall l.
\label{norm}
\end{equation}
This condition is clearly satisfied by the coefficients in Table 1. 

\section{Reconstructing the Diffraction Volume}

The average angular correlations amongst the resolution rings of the
different measured diffraction patterns contain information about
the 3D diffraction volume of a single particle. Such angular
correlations are defined by
\begin{eqnarray}
C_2(q,q',\Delta \phi) = \frac{1}{N_p} \sum_p 
\sum_{n=0}^{N-1} I_p(q,\phi_n) I_p(q',\phi_n+\Delta \phi) \nonumber \\
= \frac{1}{N_p} \sum_p \frac{1}{N} \sum_{m=0}^{N-1} I_m(q) I_m(q')^*
\exp{(-im\phi_n)}
\label{c2}
\end{eqnarray}
where $I_p$ is the intensity on diffraction pattern $p$, $N_p$ is the
number of available diffraction patterns from random orientations of the
particle,
$\phi_n$ is the $n$-th of $N$ discrete values of $\phi$, and 
\begin{equation}
I_m(q)=\sum_{n=0}^{N-1} I_p(q,\phi_n) \exp{(im\phi_n)} 
\label{eyem}
\end{equation}
the angular Fourier tranform of each resolution ring of the $p$-th 
diffraction pattern.
Indeed the fastest way to calculate the average angular
correlation $C_2(q,q',\Delta \phi)$ is to exploit the cross-correlation
theorem by performing the angular Fourier
transform (\ref{eyem}) of each individual diffraction pattern, take the
product of the Fourier transform and its complex conjugate followed by the inverse transform, and to average the results over 
the diffraction patterns ((\ref{eyem}) plus the second equality of (\ref{c2})). 

It has been shown \cite{saldin2009} that if the data from enough diffraction
patterns of randomly oriented identical particles are averaged,
\begin{equation}
C_2(q,q',\Delta \phi) = \sum_l F_l(q,q',\Delta \phi) B_l(q,q')
\label{bl_eqn}
\end{equation}
where 
\begin{equation}
F_l(q,q',\Delta \phi)= \frac{1}{4\pi} P_l[\cos{\theta(q)}\cos{\theta(q')}+
\sin{\theta(q)}\sin{\theta(q')}\cos{(\Delta \phi)}]
\end{equation}
where $P_l$ is a Legendre polynomial of order $l$,
\begin{equation}
\theta(q)=\pi/2-\sin^{-1}(q/2\kappa),
\end{equation} 
$\kappa$ is the wavenumber of the incident beam, and
\begin{equation}
B_l(q,q')=\sum_m I_{lm}^*(q) I_{lm}(q')
\label{bl}
\end{equation}   
Since in Eq.(\ref{bl_eqn}), the LHS may be found from experiment, and 
$F_l(q,q',\Delta \phi)$ is a known mathematical function, $B_l(q,q')$
may be found by solving this equation. Due to its form, $B_l(q,q')$
contains information about the 3D diffraction volume of the particle
via the spherical harmonic expansion coefficients $I_{lm}(q)$. Because
the RSHs are related to the regular spherical harmonics by a unitary
transformation, one may also write
\begin{equation}
B_l(q,q')=\sum_m R_{lm}(q) R_{lm}(q').
\label{blr}
\end{equation}
This is a more convenient form since all quantities in this equation are
real. If the $R_{lm}(q)$ coefficients may be found from deduced values
of $B_l(q,q')$, the expression (\ref{realexp}) is just as convenient
for reconstructing the 3D diffraction volume as (\ref{genexp}).

However, finding the correct $R_{lm}(q)$'s from known $B_l(q,q')$'s is
still a formidable task since it involves taking a matrix square root
\cite{saldin2009}. Such a square root is necessarily ambiguous by an
orthogonal matrix which cannot be found from the $B_l(q,q')$'s alone.
In principle, such an orthogonal matrix may be found from the so-called  
angular triple correlations \cite{kam1980} or by an iterative phasing 
algorithm that alternately satisfies constraints to the meaured 
angular correlations and in the 3D space of the reconstructed 
intensity disrtibution \cite{saldin_prb_2010}.

When the particle under study is known to have a high degree of
symmetry, like an icosahedral virus, this problem is greatly simplified.
Comparing (\ref{realexp}) and (\ref{icoexp}) with definition (\ref{icodef}),
one can deduce that, for a diffraction volume with icosahedral symmetry,
\begin{equation}
R_{lm}(q) = g_l(q) a_{lm}
\label{icos}
\end{equation} 
Substituting (\ref{icos}) into (\ref{blr}) we see that one may write
\begin{equation}
B_l(q,q') = g_l(q) g_l(q') \sum_m a_{lm}^2
\end{equation}
and using (\ref{norm}) this may be simplified further to
\begin{equation}
B_l(q,q') = g_l(q) g_l(q') 
\label{bg}
\end{equation}

We see here the great advantage of using IH's rather than RSHs for this
problem of icosahedral symmetry. The sum over $m$ in the RHS of (\ref{blr})
has disappeared completely in the RHS of (\ref{bg})! The RHS of this
equation is just the product of two scalars.
A diffraction volume of icosahedral symmetry may be reconstructed via
(\ref{icoexp}) if the coefficients $g_l(q)$ are known. Since the other
quantities in (\ref{icos}) are real, it is clear that $g_l(q)$ may be chosen to
be real. The magnitudes of the $g_l(q)$ coefficients may be found from the
diagonal quantities $B_l(q,q)$ deduced from the intensity autocorrelations
on resolution ring $q$ via
\begin{equation}
|g_l(q)|=\sqrt{B_l(q,q)}
\end{equation}
Thus, the only remaining task in determining the coefficients $g_l(q)$ is
determing the signs of these real numbers.
A simple way is to notice that the expression (\ref{icoexp}) for the 
intensity of a resolution shell of radius $q$ in the 3D diffraction volume 
may be rewritten
\begin{equation}
I(q,\theta,\phi)=\sum_l |g_l(q)| sign[g_l(q)] J_l(\theta,\phi)
\label{ico2}
\end{equation}
where the only unknown quantities in the RHS are the signs of $g_l(q)$.
Since the only permitted values of the quantum number $l$ of the icosahedral 
harmonic coefficients $g_l(q)$ of a diffraction volume are the even permitted
values up to $l$=30, namely $l$=0, 6, 10, 12, 16, 18, 20, 22, 24, 26, 28, 
and 30, we attempted to determine these signs by an exhaustive search
over the 2$^{12} \simeq 4000$ combinations of signs by finding the 
combination that minimized 
\begin{equation}
\sum_{\theta,\phi} |I_{-}(q, \theta, \phi)|
\end{equation}
where $I_{-}$ are the negative values of $I$, for a chosen resolutiuon shell 
$q$. The physical basis of this is simply that $I(q,\theta,\phi)$ has to
be a positive definite quantity, and our best approximation to this is a
function with a minimum sum of the magnitudes of negative values. As subsequent 
results show, this easily implemented prescription seemed accurate
enough to find a good enough approximation to the correct signs of these 
coefficients for the chosen reference resolution ring. To maximixe the number 
of non-negligible magnitudes $|g_l(q)|$ we chose a high-resolution 
resolution ring. In order to avoid almost all values $|g_l(q)|$ being
very small, and thus subject to significant rounding-off errors,
we found the best compromise to choose the reference ring to be one for
which $q \simeq \frac{2}{3} q_{max}$, where $q_{max}$ is the value of $q$ for
the outermost resolution shell.

From Eq. (\ref{bg}), we see that the icosahedral
harmonic expansion coefficients of the same quantum number $l$, corresponding
to a different resolution shell $q'$ are related to the now known ones
of resolution shell $q$ by the simple quotient
\begin{equation}
g_l(q')=B_l(q,q')/g_l(q),
\end{equation} 
Thus, having found the coefficients $g_l(q)$ for a paticular shell $q$,
those of the other shells $q'$ were determined from this simple quotient,
involving the quantities $B_l(q,q')$ directly calculable from the
average intensity cross correlations between different resolution rings
on the measured diffraction patterns. Thus the exhaustive search though
all 2$^{12}$ combination of signs  
needs to be performed 
only for a single resolution ring $q$. 

A knowledge of the expansion coefficients for all the resolution shells
should enable a reconstruction of the 3D diffraction volume via
(\ref{icoexp}). If this intensity distribution is interpolated onto
an oversampled \cite{miao1999} 3D Cartesian reciprocal-space grid, 
$(q_x, q_y, q_z)$, say,  
an iterative phasing algorithm \cite{marchesini} may be applied to reconstruct 
the 3D electron density of the scattering particle.

\section{Numerical Tests}

A central thesis of this paper is that the 
the scattered intensity from an icosahedral particle may be represented
by a sum of icosahedral harmonics. We first sought to verify this
proposition by calculating first the spherical harmonic expansion
coefficients of a simple icosahedral particle via the expresion
\begin{equation}
A_{lm}(q) = i^l \sum_j f_j(q) j_l(qr_j) Y_{lm}({\hat{r}}_j),
\end{equation} 
where $f_j(q)$ is the form factor of the jth atom, ${\bf r}_j$ is its
coordinate, and $j_l$ is a spherical Bessel function of order $l$. 

\begin{figure}[htbp]
\centering\includegraphics[width=7cm]{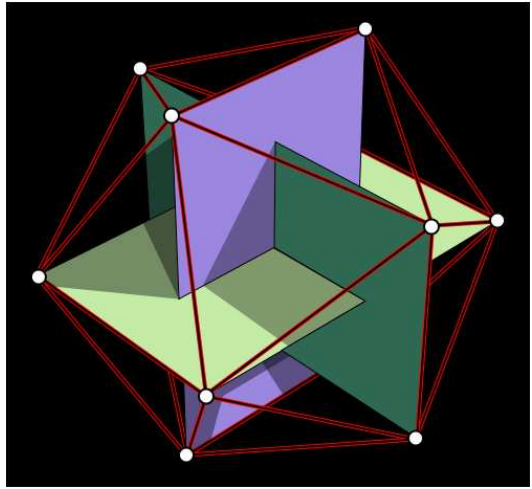}
\caption{Regular icosahedron \cite{wiki}} 
\label{regico}
\end{figure}

For our initial tests we simulated the scattering from an artificial
molecule of identical atoms at the vertices of a regular icosahedron 
(Fig. \ref{regico}) of edge
length 2 \cite{wiki} (which we take to be to be in \AA \ units, with 
Cartesian coordinates (also assumed to be in \AA):
\\ \\
(0, $\pm$1, $\pm \Phi$)
\\
($\pm$ 1, $\pm \Phi$, 0)
\\
($\pm \Phi$, 0, $\pm$ 1)
\\ \\
where $\Phi$ is the golden ratio (1+$\sqrt{5}$)/2.

\begin{figure}[htbp]
\centering\includegraphics[width=7cm]{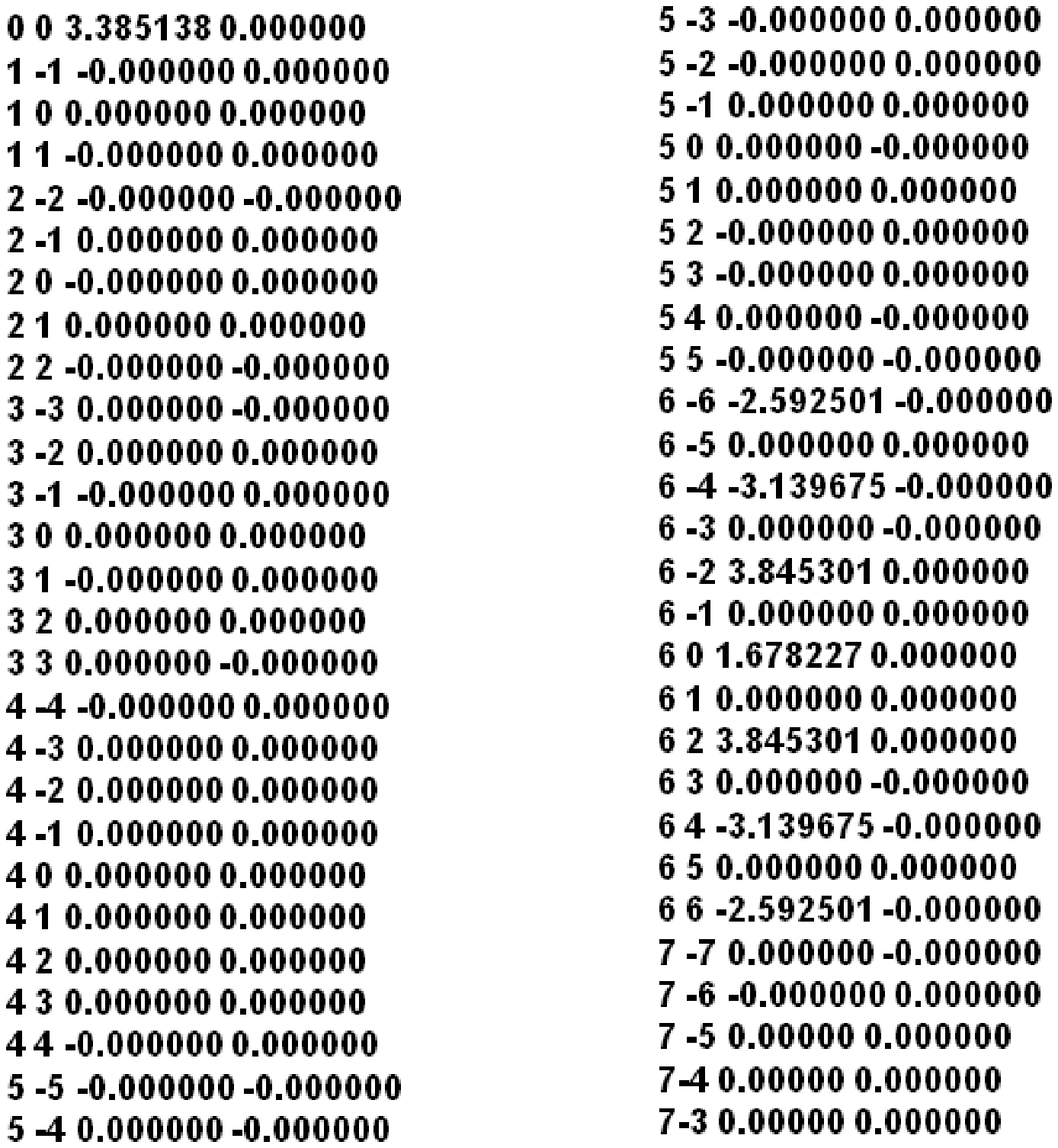}
\caption{Calculated values of the $A_{lm}$ coefficients (arbitrarily taking 
$f_j(q)=1,$ $\forall j$) assuming 12 identical atoms at the vertices of a 
regular icosahedron. The first two entries in each column in each line are 
the $l$ and $m$ values. The next two are the real and imaginary parts of 
$A_{lm}(q)$.  It will be seen that all coefficients are zero except those for 
which $l$=0 or 6, and that all non-zero coefficients are real.}
\label{alm_coeff}
\end{figure}

The resulting calculated values of the amplitudes $A_{lm}$ (arbitrarily
taking $f_j(q)=1, \forall j$) for all possible values of 
of $l$ and $m$ are listed in Fig. \ref{alm_coeff}. Note that the amplitudes 
$A_{lm}$ are all real, and that, for the values listed, they are non-zero 
only for $l$=0 and 6. The values for $l$=1,2,3,4,5, and 7 are all seen to be 
zero, corresponding to non-existing icosahedral harmonics for these values of 
$l$. Here too, all coefficients are zero except those for which $l$=0 or 6,
and all non-zero coefficients are real. Note that this result will be true
for any icosahedral orientation since the a rotation matrix (Wigner D-matrix)
will mix only amplitudes of different magnetic quantum number corresponding
to the same angular momentum $l$. The z-axis of the simple
icosahedron used for this test is a 2-fold rotation axis, not 5-fold, 
unlike e.g. Fig.  \ref{icoharm} above, or Table 1 below. 
This is why the amplitudes corresponding to
every other value of $m$ are non zero for $l$=6 rather than every
integer multiple of 5, when $z$ is chosen to be a 5-fold axis. 

\begin{figure}[htbp]
\centering\includegraphics[width=7cm]{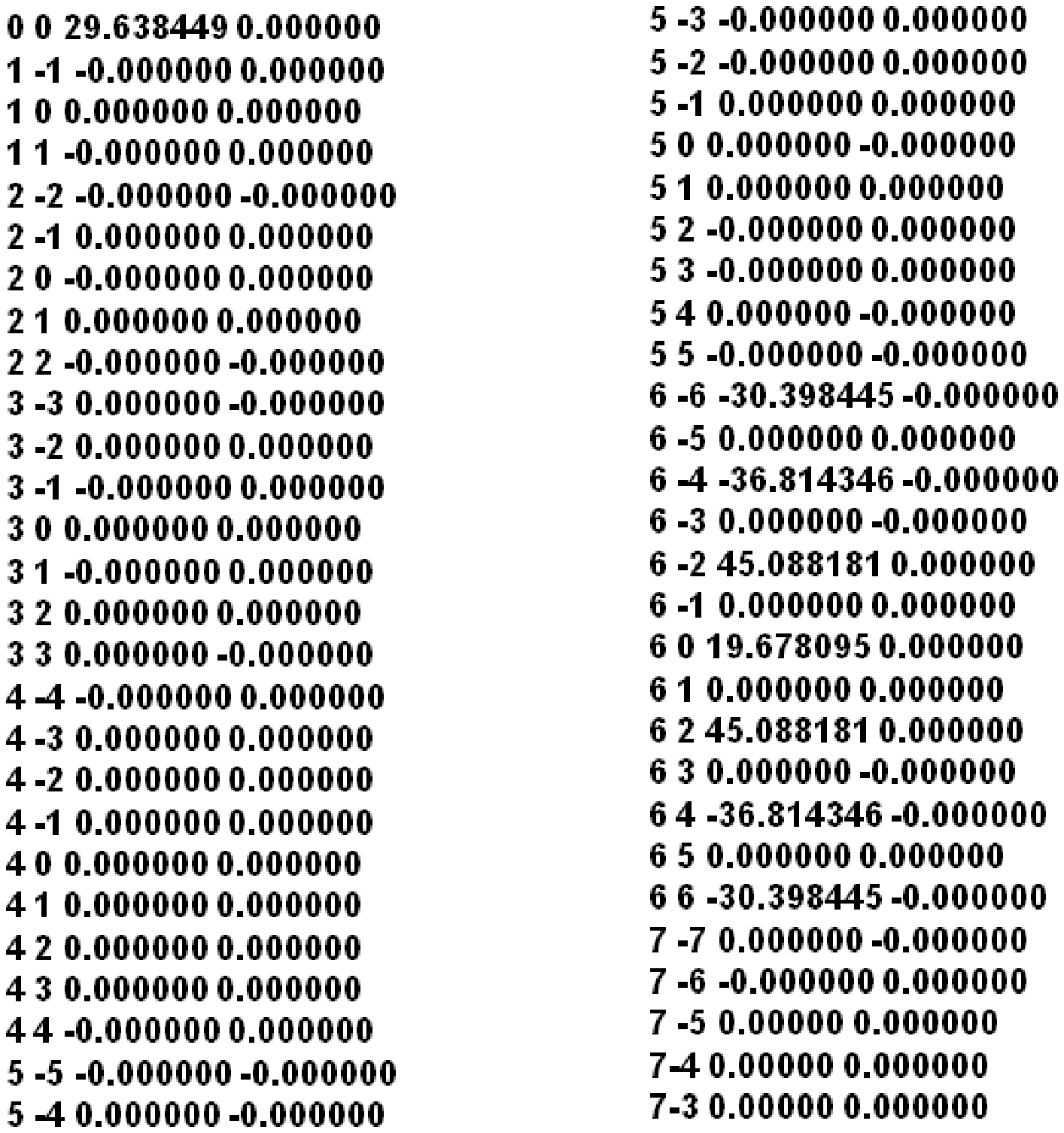}
\caption{Same as Fig. \ref{alm_coeff}, except for values of the $I_{LM}(q)$ 
coefficients calculated by Eq. (\ref{clebsch}) from the $A_{lm}(q)$ values in 
Fig. \ref{alm_coeff}.}
\label{ilm_coeff}
\end{figure}
Of greater interest for our method are the allowed values of $L$ for
the coefficients, $I_{LM}$, of the spherical harmonic expansions of the 
scattered intensity. Since 
\begin{equation}
I(\vec{q})= |A(\vec{q})|^2, 
\label{eye}
\end{equation}
it must
follow that if $A({\bf q})$ has icosahedral symmetry, so must $I({\bf q})$.
However, this is not entirely obvious from the relationship between
the two sets of coefficients
\begin{eqnarray}
I_{LM}(q) = \sum_{lm;l'm'} A_{lm}(q) A_{l'm'}^*(q) 
\int Y_{lm}({\bf \hat{q}})Y_{l'm'}^*({\bf \hat{q}})
Y_{LM}^*({\bf \hat{q}}) d{\bf \hat{q}} \nonumber \\
=\sum_{lm;l'm'} A_{lm}(q) A_{l'm'}^*(q)
\int Y_{lm}^*({\bf \hat{q}})Y_{l'm'}({\bf \hat{q}})
Y_{LM}({\bf \hat{q}}) d{\bf \hat{q}} \nonumber \\ 
=\sum_{lm;l'm'} A_{lm}(q) A_{l'm'}^*(q)
\sqrt{\frac{(2l'+1)(2L+1)}{4\pi(2l+1)}}
C^{l0}_{l'0L0} C^{lm}_{l'm'LM}
\label{clebsch}
\end{eqnarray}
where $C^{lm}_{l'm'LM}$ is a Clebsch-Gordan coefficient 
\cite{varshalovich1989}.
According to the usual theory of the vector addition of angular momenta,
the allowed values of $L$ are all integers in the range from 
$|l-l'|$ to $l+l'$, with no obvious indication that $L$=1,2,3,4,5, and 7,
for instance, are forbidden. However, a straightforward evaluation of
the $I_{LM}(q)$ coefficients via (\ref{clebsch}) reveals this to be the
case, as is seen by the tabulated values of these coefficients in 
Fig. \ref{ilm_coeff}. 

\begin{figure}[htbp]
\centering\includegraphics[width=7cm]{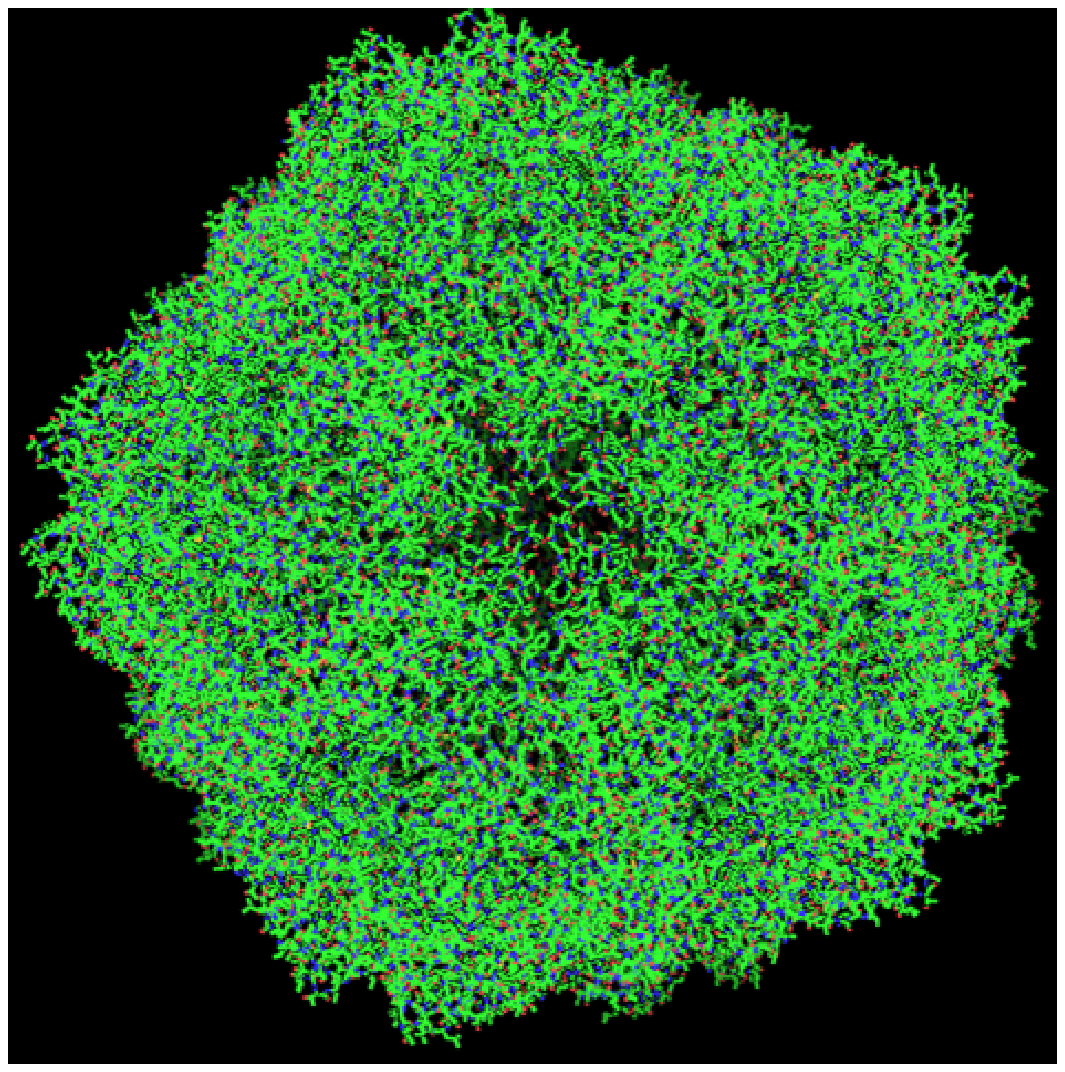}
\caption{Top part of the structure of the satellite tobacco necrosis virus
(STNV) viewed down its 5-fold axis (from structure data in 
PDB entry: 2BUK)}.
\label{stnv}
\end{figure}

\begin{figure}[htbp]
\centering\includegraphics[width=7cm]{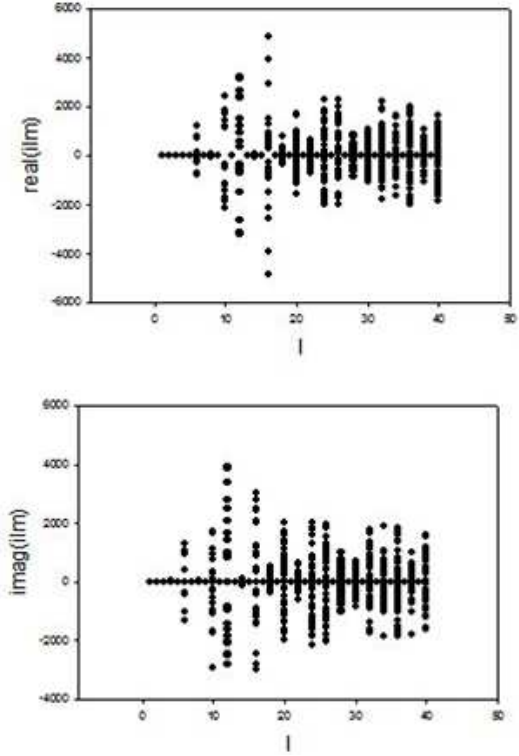}
\caption{Real and imaginary parts of the $I_{lm}(q)$ coefficients
calculated from the computed diffraction volume of STNV. Each dot 
represents a value of the (lm) pair. Note that these coefficients
are largely absent for $l$=1,2,3,4,5,7,8,9,11,13,14,15.} 
\label{ilm}
\end{figure}

We next tested this on a realistic model of the small
icosahedral virus, satellite tobacco necrosis virus (STNV) whose
atomic coordinates are deposited in the protein data bank under entry
2BUK (Fig. \ref{stnv}). We calculated $A({\bf q})$ from the usual structure 
factor expression
\begin{equation}
A(\vec{q})=\sum_j f_j(q) \exp{(i\vec{q}\cdot\vec{r}_j)}
\end{equation}
and constructed the diffraction volume from (\ref{eye}). By integrating over 
spherical shells of $I(\vec{q})$ we evaluated the spherical harmonic
expansion coefficients of the 3D diffraction volume of STNV from
\begin{equation}
I_{lm}(q)= \int I(\vec{q}) Y_{lm}({\bf \hat{q}}) d{\bf \hat{q}},
\end{equation} 
where $\hat{q}$ is the unit vector $\vec{q}/q$, 
with this integration conveniently performed by Gaussian qudrature
\cite{rabinowitz}. Plots of the real and imaginary parts of $I_{lm}$
in Fig. \ref{ilm} clearly show the same trend of vanishing components
corresponding to $l=1,2,3,4,5,$ and 7 and in addition vanishing
components for $l=8,9,11,13,14,$ and 15, exactly consitent with the
tablulated values of icosahedral expansion coefficients in Table 1. 
What is more, it was found that 
\begin{equation}
(-1)^m I_{l(-m)}(q)=I_{lm}(q),
\end{equation}
the precise condition for the reality of the $R_{lm}(q)$ coefficients
of the RSHs, and hence of the icosahedral harmonic expansion coefficients
$g_l(q)$ via (\ref{icos}).

Since this result is a consequence of the icosahedral symmetry of
the diffraction volume $I({\bf q})$, it is to be expected of the
diffraction volume of all icosahedral viruses (assuming the
protein coat to be the dominant scatterer). In view of (\ref{bl})
this must mean that the $B_l(q,q')$ coefficients computed from the
data of diffraction patterns of random orientations of all icosahedral
particles must all have vanishing values for 
$l=1,2,3,4,5,7,8,9,11,13,14,15,..$, thus providing a very simple test
of whether the diffraction patterns measured in a ``diffract and destroy''
experiment with a FEL are from an icosahedral particle.

Assuming this is indeed found to be approximately true in practice (even 
the so-called icosahedral viruses may have appendages which break the 
icosahedral symmetry of the protein coat, and of course the genetic material 
inside the protein coat would not be expected to have this symmetry.
However, if the bulk of the material of the virus may be assumed to constitute
the protein coat, this must be approximately the case). The icosahedral
structure of the protein coat may be found by an analysis of the large
$l=0,6,10,12,16,...$ $B_l(q,q')$ coefficients extactable from the average
angular correlations of the diffraction data.

\section{Reconstruction of STNV from Simulated Diffraction Patterns}

We next attemped a reconstruction of satellite tobacco necrosis virus (STNV)
from diffraction patterns simulated for directions of incidence on a
single particle from a uniform angular distribution in SO(3) 
\cite{lovisolo}. For the model of STNV we took the data of the bological 
assembly of STNV from PDB 
entry 2BUK. Due to the large number of atoms in this biological assembly
($\sim$ 100,000), the most convenient way to do this was to take 
slices through a precalculated 3D diffraction volume of this structure. 
Average angular correlations of these simulated diffraction patterns were 
calculated by the formulae (\ref{c2}) and (\ref{eyem}). and the $B_l(q,q')$ 
coefficients were calculated from these by inverting Eq. (\ref{bl_eqn}). 

For the 10,000 simulated diffraction patterns in our test, this process took 
about a quarter of an hour on a single processor on a desktop computer. In a 
real experiment, one may 
have to deal with perhaps 100 times as many diffraction patterns, with more 
pixels per pattern, so the processing time could be several orders of magnitude
greater. However, the bulk of the time will be spent in generating
the average angular correlations $C_2(q,q',\Delta \phi)$ (\ref{c2}),
a process which easily lends itself to parallelization, since subsets of
the diffraction patterns may be averaged by separate computer 
processors, and the averages themselves subsequently averaged. Nevertheless, 
this process of reduction of terabytes (TB) of measured experimental 
data is probably the most computer-resource intensive part of our
method. Having thus reduced our data to a set of $B_l(q,q')$ coefficients
for a set of 30 values of $l$, and 61 values of $q$ (and $q'$), we were
left with a set of 30$\times$61$\times$61 real numbers which formed  
the input to our reconstruction algorithm. This required about a MB
of storage/memory. In a real experiment also, our method requires the
million-fold reduction of the TB of data to a MB of 
floating-point (real) numbers that form the input to our reconstruction
algorithm. It is recommended that this data reduction be performed at
the site of the data to reduce by a million-fold or so the quantity of
data that needs to be transmitted over the internet to the site
where the image reconstruction is performed. At current rates, the transmission
of Terabytes of data over the internet could take several weeks, whereas
the time for the transmission  of a MB of data could be measured in
seconds. In addition this process of data reduction is expected to result
in considerable noise-reduction of the raw data though averaging 
\cite{poon_2011}. 

Since the $B_l(q,q')$ coefficients are related to the expansion coefficients
$R_{lm}(q)$ of the real spherical harmonics (which satisfy the same
selection rule on $l$ as do the expansion coefficients $I_{lm}(q)$ of
the regular spherical harmonics), it would be expected that the
$l=0,6,10,12,16,18,20,22,24,26,28,30$ elements of these coefficients are 
dominant. This was found to be the case for our simulations of STNV. Some 
of the larger, predominantly icosahedral, viruses may have appendages like 
the unique vertex and ``hair'' of the mimivirus \cite{MGR_mimi} , or the
spike fom a unique vertex of the chlorella virus \cite{chlorella}.  Indeed, 
with values of these coefficients extracted from experimental single-particle
diffraction patterns from an unknown particle, the satisfaction of
this selection rule would be an excellent test of the degree to which
the particle is icosahedral. Inclusion of only the large
$l=0,6,10,12,16,...$ of the $B_l(q,q')$ coefficients in the reconstruction
algorithm consistent with icosahedral symmetery is equivalent to finding the 
closest icosahedral approximation to the structure.

\begin{figure}[htbp]
\centering\includegraphics[width=7cm]{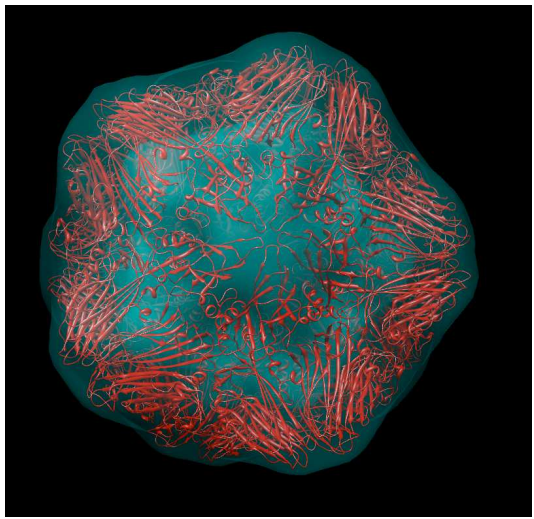}
\caption{Reconstructed image from the diffraction volume of a single STNV 
particle computed directly from a structure factor calculation. 
STNV is about 20 nm in diameter. The figure depicts a view of the
icosahedron close down its 5-fold rotation axis. The reconstruction 
assumed a maximum value of $q$, $q_{max}$, of about 4.7 nm$^{-1}$,
implying a resolution of $\sim$~1.3 nm. Both the outer and inner
surfaces of the virus capsid are apparent in this representation.
A ribbon diagram of the structure in PDB entry 2BUK is seen to fit
within this capsid.}
\label{direct}
\end{figure}

The procedure described in section 3 was then followed to reconstruct a 3D 
diffraction volume, consisting of set of scattered intensities 
$I(q_x,q_y,q_z)$ over 3D reciprocal space as a function of the
reciprocal-space coordinate $\vec{q}\equiv(q_x,q_y,q_z)$. In our simulations,
we took this to be a 61$\times$61$\times$61 array of real numbers.
The computer time for this process was almost ridiculously short, amounting to
no more than a few seconds on a single-processor desktop computer. 

\begin{figure}[htbp]
\centering\includegraphics[width=7cm]{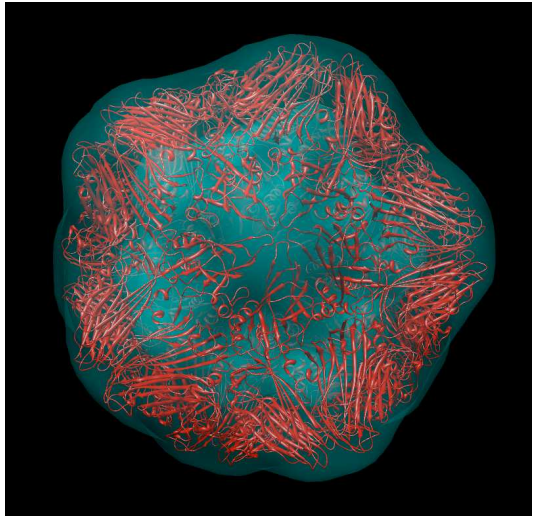}
\caption{
Same as Fig. \ref{direct} except that the diffraction volume was 
reconstucted from the average of angular correlations on 10,000
diffraction patterns of STNV from uniformly distributed directions
over SO(3). The reconstructed electron density is seen to be remarkably 
similar to that in Fig. \ref{direct}.}
\label{reconstr}
\end{figure}

\begin{figure}[htbp]
\centering\includegraphics[width=7cm]{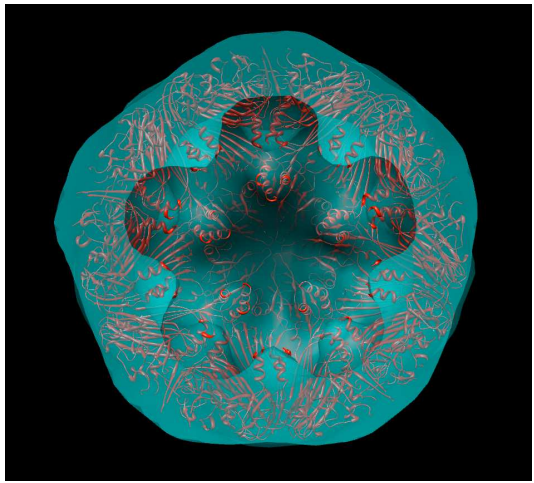}
\caption{
Same as Fig. \ref{reconstr} except that image displayed is a cut
perpendicular to the 5-fold axis of the virus. The 5-fold symmetry of
both the external and internal surfaces of the capsid in this
projection are clearly visible.}
\label{cutout}
\end{figure}

The final step is the recovery of a 3D electron density of the particle.
This may be done by a standard iterative phasing algorithm. We used the
``charge flipping'' algorithm of Oszl\'{a}nyi and S\"{u}to \cite{os1,os2}. 
In order to judge the accuracy of the our algorithm in recovering the 3D 
diffraction volume, we performed this recovery of the 3D electron 
density from both the diffraction volume $I({\bf q})$ calculated directly 
from the STNV structure factors (Fig. \ref{direct}), and also by our algorithm 
from the $B_l(q,q')$ coefficients (Fig. \ref{reconstr}), which may be computed 
from the measured data of the FEL diffraction patterns from random particle 
orientations. The similarity of the reconstructed images of Figs.
\ref{direct} and \ref{reconstr} was a further indication of the validity of 
the method of image reconstruction
from the quantities $B_l(q,q')$ derivable from the average angular 
correlations. The fact that the reconstructed image consists of a thin
protein shell is also seen from the slice perpendicular to a 
5-fold rotation axis through the reconstructed image of Fig. \ref{reconstr}
depicted in Fig. \ref{cutout}. In the case of all three figures, a ribbon
representation of the structure from the biological assembly of the STNV
virus from the same PDB structure used to simulate the diffraction patterns
is superimposed on the semi-transparent electron density to show the 
excellence of the reconstruction. 
It should be emphasized that nowhere in our theory is
it assumed that the structure consists of a thin protein shell, unlike the
so-called shell model that has been used in the SAXS analysis of virus
capsids \cite{zheng1995}. In our case, the existence of a shell is
deduced by an iterative phasing algorithm from the anaysis of data 
from diffraction patterns of random particle orientations without
any assumptions on our part.

\section{Beyond the Icosahedral Approximation}

Satellite tobacco necrosis virus (STNV) is an example of a virus with 
a perfectly icosahedral protein coat \cite{viperdb}. A host cell gets access
to the genetic material of this virus by ingesting it whole and dissolving 
its protein coat.

Many of the larger viruses are only approximately icosahedral: they often have
appendages, such as a neck sticking out of the coat that is used to inject 
the genetic material inside the coat into a host cell whose protein 
making capability is hijacked by the virus DNA or RNA.

An ultimate reconstruction algorithm should be able to reconstruct these
non-icosahedral parts of the structure in addition to the icosahedral part.
The above procedure has determined the icosahedral harmonic expansion
coefficients $g_l(q)$ that best fit the measured quantities
$B_l(q,q')$. Any deviations from these values are due to
the non-icosahedral parts of the structure. Any differences between
the experimental values of $B_l(q,q')$ and 
$g_l(q)g_l(q')$ may be written
\begin{equation}
\delta B_l(q,q')=\sum_m a_{lm} \{ g_{l}(q)  \delta R_{lm}(q')
+ \delta R_{lm}(q) g_{l}(q') \} + \delta R_{lm}(q) \delta R_{lm}(q'), 
\label{delta_b}
\end{equation}
in terms of $\delta R_{lm}(q)$, the extra contribution to the RSH
expansion coefficients due to deviations from icosahedral symmetry.
Note that for $(l,m)$ combinations not associated with icosahedral
harmonics, e.g. those for which there is no entry in a list like 
Table 1, the terms $a_{lm}$ will be zero, and only the 
quadratic terms in $\delta R_{lm}$ will survive in (\ref{delta_b}).
Determination of the $\delta R_{lm}(q)$ coefficients which optimize
the agreement the theoretical expression (\ref{delta_b}) and the measured
values will enable the construction of a better estimate of
a single-particle diffraction volume via
\begin{equation}
I(\vec{q}) = \sum_{lm} \{ g_l(q) a_{lm} + \delta R_{lm}(q) \} 
S_{lm}({\bf \hat{q}}).
\end{equation}
The presence of the correction terms $\delta R_{lm}(q)$, which
have no symmetry restrictions (apart from Friedel symmetry) will allow
the diffraction volume calculated by this formula to include 
deviations from icosahedral symmetry.

Application of an interative phasing algorithm to an oversampled diffraction
volume calculated by this expression will enable the determination
of the full structure of the virus, including any appendages that break the 
approximate icosahedral symmetry.
 
\section{Discussion}

The remarkable similarity of the reconstructed electron densities of
Figs. \ref{direct} and \ref{reconstr}, and the fit of the latter to the model 
of STNV from the PDB file, are indications of the correctness of the method 
of reconstruction of the 
3D diffraction volume from the average angular correlations of the
10,000 simulated diffraction patterns of STNV. We calculated from
these the $B_l(q,q')$ coefficients for all values of $l$ from 0 to $30$. We 
found good agreement with the selection rule on the $l$ coefficients
in which the sizes of the $B_{q,q'}$ coefficients for all odd values
of $l$ were small (due to Friedel, or inversion, symmetry) and in addition the
even values $l$=2,4,8, and 14 were also small, due to the icosahedral symmetry
of the 3D diffraction volume of a single particle. We included $g_l(q)$ 
coefficients for the non-negilible $B_l(q,q')$ coefficients up to 
$l$=30 (up to which value the icosahedral harmonic expansion coefficients
depend on the $l$ quantum number only). If $q_{max}$ is the maximum
value of the reciprocal-space coordinate $q$ up to which the reconstruction is
valid, conventional wisdom \cite{pendry1974} suggests that $l_{max}$ and 
$q_{max}$ should be related by 
\begin{equation}  
q_{max} R = l_{max},
\label{limit}
\end{equation}
where $R$ is the radius of the particle. Taking
\begin{equation}
q_{max}=2\pi/d
\label{qmax}
\end{equation}
where $d$ is the resolution. Substituting (\ref{qmax}) into (\ref{limit})
and rearranging, we find that
\begin{equation}
d/R=2\pi/l_{max} \simeq 1/5.
\end{equation}
STNV has a radius of $\sim$ 100 \AA \ suggesting a resolution of about 20 
\AA. In practice we found that increasing $q_{max}$ a further 50\% or so,
while keeping $l_{max}$ fixed at 30 seemed to improve the quality of the
reconstructed image. Presumably because up to about 1.5$q_{max}$ the 
spherical harmonic expansion coefficients of $l$ greater than 30 remain
small. 

It should be emphasized this is not necessarily an absolute limit of
the resolution obtainable with the use of icosahedral harmonics. The
higher order harmonics, at least up to $l$=44, have been tabulated
by Zheng {\it et al.} \cite{zheng1995}. At least up to this value, the 
degeneracy of the icosahedral harmonics characterized by a particular
value of $l$ is no more than two. Although the algorithm for recovering
the expansion coefficients of such degenerate hamonics from the
experimental data is a little more complicated, it seems far from
an insuperable problem.

The images in Figs. \ref{direct} to \ref{reconstr} were computed by an
iterative phasing algorithm \cite{os1,os2} from a reciprocal-space 
distribution of intensities oversampled \cite{miao1999} by a factor of 
$\sim$ 2 with respect to the size of STNV, up to a
$q_{max}$ value of $\sim$ 0.47 \AA$^{-1}$ (a 61$\times$61$\times$61 array), 
implying a resolution of about 13 \AA, and a $d/R$ ratio closer to
$1/8$. Further, the images of Fig. \ref{direct}-\ref{cutout} reveal 
this coat to be hollow. 
The slice (Fig. \ref{cutout}) through the reconstructed image
perpendicular to the 5-fold axis reveals both external and internal
surfaces of 5-fold rotational symmetry. The revelation of the 
hollow nature of the protein coat is of course an extra feature
contained in the 3D intensity distribution above and beyond the assumed 
icosahedral symmetry. It is revealed by the iterative reconstruction
algorithm used \cite{os1,os2} due to the paricular variation of the $B_l(q,q')$
coefficients with the radial reciprocal-space coordinates $q$ and $q'$. 

Some the advantages of this method of analysis of single particle 
diffraction patterns from unknown particle orientations compared with 
other proposed algorithms \cite{fung2009,loh2009} should be pointed 
out. Since it has been shown 
\cite{kam1977,saldin_njp_2010,saldin_prb_2010,poon_2011,saldin_prl_2011}, 
that the angular correlations of multiple identical particles in arbitrary 
orientations are essentially identical to those from a single particle,
the method we have described is equally applicable to droplets
containing multiple particles injected into the XFEL \cite{deponte}
as to the injection of single particles in random orientations. Thus there is
no need to discard diffraction patterns from multiple particle hits.

Since the inputs to our algorithm are not the direct
photon counts, but rather the average of the angular correlations between
intensities of 
the {\it same} diffraction patterns, it is insensitive to shot-to-shot
fluctuations between the diffraction patterns, as may be caused by
intensity variations of the incident X-ray beam or, for example, by the
number of particles scattering a particular X-ray pulse.

The raw experimental data is likely to consist of $\sim$ 10$^6$ diffraction 
patterns, each of $\sim$ 10$^6$ pixels. Thus the raw experimental data
will require TB of storage. Of course, this is very noisy data, and
the structural information content is much less than this. The averaging
of the angular correlations that we perform may be regarded as a form
of data averaging that results in infomormation concentration and
noise reduction. Even if the 
number of values of q chosen is, say, 61, and these coefficeints are 
evaluated for, say, 30 values of $l$, the total number of these (real) 
coefficients will be only of the order of 100,000, requiring less than a MB 
of storage. This data reduction is best performed at the site of the data
to allow the tranference of a million times less data over the
internet to the site of image reconstruction. The reconstruction of
3D images of the quality of Figs. 8-10 from a properly constructed
set of $B_l(q,q')$ coefficients is extraordinarily rapid. In our calculations,
reconstruction of an array of $I(q_x,q_y,q_z)$ values representing a 3D 
diffraction volume at reciprocal-space coordinates 
$\vec{q}\equiv(q_x,q_y,q_z)$ on a 61$\times$61$\times$61 Cartesian
grid took just a few seconds on a single Intel Q6600 processor, using an 
Intel Fortran compiler. The reconstruction of real-space images 
of the quality of Figs. 8-10 from this array by means of a ``charge
flipping'' algorithm \cite{os1,os2} took a further 4 minutes for 200 
iterations on a laptop PC.  

Of course, the averaging of the data from the different diffraction
pattterns assumes they all arise from copies of the particle in different
orientations (as does the technique of small angle X-ray scattering,
SAXS, for example). In order to distinguish between different conformations 
of the individual molecules, it may be necessary to 
operate on the entire ensemble of all the measured diffraction patterns 
One of the disadvantages of such methods is the need to operate on 
single-particle diffraction patterns and thus, unlike with 
our method, diffraction patterns from multiple hits need to be removed. Such 
methods also face the problem of the uncertain normalization of incident 
intensities between successive pulses of incident radiation. Also, in
contrast to our method, such techniques may require the tranferance of 
perhaps TB of data over the internet to the site of the data 
performing the analysis, which needs to be equipped with a cluster of 
computers performing parallel computations.  

\section{Conclusions}

When reconstructing the structure of a virus  
from ``diffract and destroy'' type single-particle diffraction experiments 
proposed for the free electron laser \cite{neutze2000}, one may exploit the 
dictum of Caspar and Klug \cite{caspar1962}
that ``there are only a limited number of efficient 
designs possible for a biological container which can be constructed 
from a large number of identical protein molecules, The two basic designs
are helical tubes and icosahedral shells''. We offer here a 
solution for the case of icosahedral viruses. For those viruses which
are substantially, though not completely icosahedral, the method proposed
is expected to be useful nontheless for initially reconstructing
the approximate icosahedral structure. The deviations from this
structure can then be found by a perturbation theory which does not
impose this symmetry.
  
The input to the algorithm is data from diffraction patterns
of randomly oriented identical particles (where the particle orientations
are unknown) in the form of the average of the angular correlations. 
As a consequence of any approximate icosahedral symmetry of the scattering 
particles, the angular
momentum decomposition of the angular correlations contains only a 
few dominant contibutions from low values of the angular momenta. 

This immediately suggests a simple test of whether the experimentally 
measured data are from the scattering by an icosahedral object. If so,
the components of the quantities $B_l(q,q')$, derivable from the
angular correlations, should have much smaller values for 
$l$=1,2,3,4, and 5 than for $l$=0 and $l$=6, for example. What is more,
as we have shown in this paper, the coefficients $g_l(q)$ of
the icosahedral harmonic expansion of the 3D diffraction volume of
the particle may be derived from the $B_l(q,q')$ data and a positivity
condition on the intensities of the 3D diffraction volume. This will be
the case even if the individual diffraction patterns are a
result of scattering from more than one particle, so there will be no need to
discard the diffraction patterns from multiple particles.

Having obtained the coefficients of an icosahedral harmonic expansion,
the 3D diffraction volume may be reconstructed as a sum over
these icosahedral harmonics. By definition, the resulting diffraction
volume will have icosahedral symmetry. If this is constructed at a grid that
is oversampled by a factor of 2 in each dimension, we have shown that
a ``charge flipping'' algorithm with no fixed support contraint is able
to reconstruct a 3D image of the particle. We find that this procedure
not only reconstructs an icosahedral shape for the particle, in simulations
for the satellite tobacco necrosis virus (STNV) it even reveals the hollow
nature of the protein coat.

{We ackowledge helpful discussions with Profs. Abbas Ourmazd and John Spence,
and financial support from DOE grant No. DE-SC0002141.}

\end{document}